\documentclass[a4paper,twoside]{article}

\usepackage{epsfig}
\usepackage{graphicx}
\usepackage{subfigure}
\usepackage{subcaption}
\usepackage{calc}
\usepackage{amssymb}
\usepackage{amstext}
\usepackage{amsmath}
\usepackage{amsthm}
\usepackage{multicol}
\usepackage{pslatex}
\usepackage{apalike}
\usepackage{algorithm2e}
\usepackage[bottom]{footmisc}
\usepackage{SCITEPRESS}     
\usepackage{hyperref}
\usepackage{float}
\usepackage{booktabs}

\begin{document}

\title{Modeling iPSC-derived Endothelial Cell Transition in Tumor Angiogenesis using Petri Nets}

\author{\authorname{Adéla Šterberová\sup{1}, Andreea Dincu\sup{1}, Stijn Oudshoorn\sup{1}, Vincent van Duinen\sup{2} and Lu Cao\sup{1}}
\affiliation{\sup{1}Leiden Insisute of Advanced Computer Science, Leiden University, Leiden, The Netherlands}
\affiliation{\sup{2}Leiden University Medical Center, Leiden, The Netherlands}
\email{l.cao@liacs.leidenuniv.nl}
}


\abstract{Tumor angiogenesis concerns the development of new blood vessels supplying the necessary nutrients for the further development of existing tumor cells. The entire process is complex, involving the production and consumption of chemicals, endothelial cell transitions as well as cell interactions, divisions, and migrations. Microfluidic cell culture platform has been used to study angiogenesis of endothelial cells derived from human induced pluripotent stem cells (iPSC-ECs) for a physiological relevant micro-environment. In this paper, we elaborate on how to define a pipeline for simulating the transformation and process that an iPSC-derived endothelial cell goes through in this biological scenario. We leverage the robustness and simplicity of Petri nets for modeling the cell transformation and associated constraints. The environmental and spacial factors are added using custom 2-dimensional grids. Although the pipeline does not capture the entire complexity of tumor angiogenesis, we are able to capture the essence of endothelial cell transitions in tumor angiogenesis using this conceptually simplified solution.}

\onecolumn \maketitle \normalsize \setcounter{footnote}{0} \vfill

\section{\uppercase{Introduction}}
\label{Intro}
Angiogenesis is the process of the development of new blood vessels. Microvascular processes in angiogenesis are found to play an important role in kidney diseases, and are even described as `the base of the iceberg' for this category of disease \cite{iceberg}. The formation of blood vessels in tumors has also been shown to influence metastasis and growth rates \cite{FOLKMAN200215,cancer_angiogenesis,VEGF_cancer}.\\
In recent years, human induced pluripotent stem cells derived endothelial cells (iPSC-ECs) became influential for disease modeling, drug discovery and regenerative therapy. Microfluidic cell culture platforms have been introduced to study angiogenesis of iPSC-ECs in a physiological relevant cellular micro-environment with controlled perfusion and gradients \cite{in_vitro_angiogenesis_gradient}. In this paper, we concentrated on modeling angiogenesis induced by VEGF which is released by the hypoxic tumor cells in a microfluidic environment.\\

\section{\uppercase{Related Work}}
There are a number of mathematical and computational models developed to study different aspects of angiogenesis \cite{article_review}. A discrete mathematical model is developed for the dynamics of vascular endothelial cells in angiogenic morphogenesis \cite{doi:10.1137/15M1038773} It incorporates cell-mixing behavior and temporal length generating behavior of the blood vessel. A new mathematical model is designed to reproduce the tumour-induced vascular networks undergoing stages of growth, regression and regrowth \cite{doi:10.1098/rsif.2016.0918}. The model is able to capture capillaries at full scale and the dynamics of vessel networks at long time scales. A mathematical formalism is developed to simulate the early stages of angiogenesis based on a 3D in vitro model \cite{Bookholt2016}. The model takes into account the dynamic interaction and interchange of different phenotypes of endothelial cells and several proteins playing a role in the interaction. A hybrid model is developed to realize \textit{silico} experiments for tumor growth and angiogenesis \cite{model_paper}. This model treats each cell as an agent, incorporates phenotypic transitions of each tumor and endothelial cell and allows VEGF and nutrient fields to impact the dynamics. \\
These mathematical models are highly powerful but they are not standardized enough for comparison. It also further hampers the reproducibility. We are looking for a unified and versatile framework for modeling biological systems. Petri nets is a graphical and mathematical formalism for the modeling and analysis of concurrent, asynchronous, distributed systems \cite{10.1093/bib/bbm029}. It is shown to be a promising mathematical tool to describe and study the relationships and interactions between various parts of a biological systems e.g. metabolic pathways, organelles, cells, and organisms \cite{Carvalho2018,Valentim2022,10.1093/bib/bbx150}. Furthermore, multiple variants of the initial formalism were created (e.g. stochastic, timed, hybrid, coloured) to enable analysing dynamical properties of complex processes, from either a qualitative or a quantitative point of view \cite{CHAOUIYA2008165}. \\
In this paper, we utilized Petri nets to simulate the transitions that iPSC-ECs undergo in the tumor angiogenesis process in a microfluidic cell culture platform. A complex pipeline is designed to model the diffusion and consumption of the VEGF as well as the migration of the cells towards the VEGF gradient. \\

\section{\uppercase{Biological Details}}
\begin{figure}
    \centering
    \includegraphics[width=0.4\textwidth]{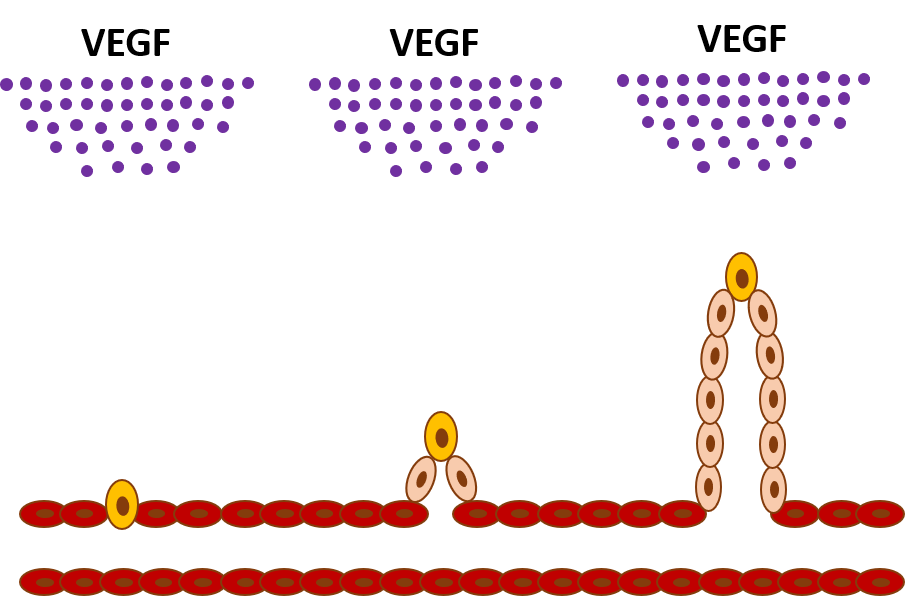}
    \caption{A visualization of the sprouting process. Red cells are phalanx cells. Brown cells are tip cells and light brown cells are stalk cells.}
    \label{fig:sprouting}
\end{figure}

A tumor induces the growth in quiescent arteries when the tumor cells are depleted of nutrients, such as oxygen. This growth induction is thus triggered in hypoxic cells, causing them to excrete VEGF. VEGF diffuses to the environment around the tumor and causes the phalanx. Endothelial cells, that make up the artery, start turning into tip cells by the influence of VEGF. In addition, the VEGF acts as a chemoattractant to tip cells. These cells inhibit neighboring cells from turning into tip cells through notch signalling \cite{notch}. These neighboring cells turn into stalk cells instead. Stalk cells proliferate and, meanwhile, follow the tip cell that is near them. The tip cells inhibit the stalk cells from transitioning back into phalanx cells within a certain distance. Therefore, the stalk cells can turn into phalanx cells to stabilise the growing artery structure. In Figure~\ref{fig:sprouting}, an example of this stage in angiogenesis is shown.

New physiologically relevant \textit{in vitro} screening was recently developed \cite{in_vitro_angiogenesis_gradient} using iPSC-ECs in a microfluidic cell culture platform, to model angiogenesis towards a VEGF gradient. This could be seen as a simplification of vessel growth near tumors, where the hypoxic tumor cells are replaced by a VEGF gradient. In this paper, the vessel formation is modeled \textit{in silico}. The aim is to create a model which includes transitions, movement, differentiation and VEGF consumption of the cell. In addition, the formation of a new blood vessel is also induced. In this process, there are several cell types with different roles. The methods that were used to simulate these are further described in the following sections.

\section{\uppercase{Method}}
\label{modeling_decisions}
We developed a hybrid model that includes a central Petri net and two grid matrices. The central Petri net realizes the cell type transitions. The two grid matrices deal in part with spatially connected features, such as the cell positioning and movement, and the growth vector (i.e. VEGF) concentrations.

\subsection{Endothelial Cell Transitions Model} \label{sec:dissecting_transition_model}
\begin{figure}[!h]
    \centering
    \includegraphics[width=0.5\textwidth]{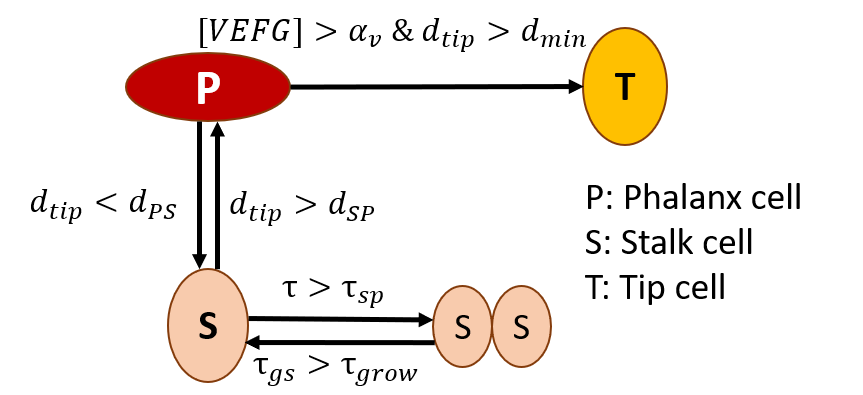}
    \caption{Schematic illustration of endothelial cell transitions. $d_{tip}$: the distance from the current cell to the nearest tip cell. $d_{SP}$: Minimum distance from tip cell for $S \rightarrow P$ transition. $d_{PS}$: Maximum distance from tip cell for $P \rightarrow S$ transition. $d_{min}$: Minimum distance from tip cell for $P \rightarrow T$ transition. [VEGF]: VEGF concentration. $\alpha_v$: VEGF threshold for $P \rightarrow T$ transition. $\tau$: the amount of time that a cell spends in a given place. $\tau_{sp}$: a predefined time interval for a stalk cell to divide. $\tau_{gs}$: growing time of a stalk cell. $\tau_{grow}$: mandatory growing time.}
    \label{fig:transitions_diagram}
\end{figure}

The endothelial cell type transitions are modeled using a timed hybrid Petri net in order to incorporate division/growing time, distance to the closest tip cell and the VEGF concentration for each cell. The scheme that we use is originally from \cite{model_paper} as illustrated in Figure \ref{fig:transitions_diagram}. Because this scheme is designed to mimic the movement of the tip cell according to the gradient of VEGF which fits our microfluidic environment the best. In addition, the paper provides a detailed baseline parameters to work with. It helps to bring biologically relevant properties to our model and makes our model more realistic.
The places in our model represent all the possible cell types that an endothelial cell can adopt, namely: phalanx cell \textit{P}, stalk cell \textit{S}, and tip cell \textit{T}. 

An additional place is added to represent the division of stalk cells ($SS$). As the effect of this transition is the separation of one cell into two smaller cells, two different tokens have to be generated, adjacent to each other (positions $(i,j)$ and $(i-1,j)$). The introduction of another token in the system has to be captured also in the positioning grid, by filling an additional position ($(i-1,j)$). A stalk cell can divide after a predefined time interval ($\tau_{sp}$), with each being half of its original size and volume. For the newly divided stalk cells to be considered again as integral stalk cells, they have to go through a growing period to increase their volume. This mandatory growing time is defined as a pre-set threshold, $\tau_{grow}$. Even if we have uniformed the volume of the cell to a constant, the mechanisms behind the growing phase are preserved.

The transitions in our Petri net correspond to possible changes in phenotype that a cell with a given type can reach.

Each transition has to be conditioned in different manners, as shown in Figure \ref{fig:transitions_diagram}. The division of a stalk cell ($S \rightarrow SS$) and the transition of divided parts back to an integral stalk cell ($SS \rightarrow S$) are conditioned by the passing of specific time intervals, $\tau_{sp}$ and $\tau_{grow}$ respectively. Additionally, the $S \rightarrow SS$ transition is constrained by the space available in the positioning grid. Both the transition from phalanx cell $P$ to stalk cell $S$ ($P \rightarrow S$) and the reversed one ($S \rightarrow P$) are conditioned by the distance of a given cell to the nearest tip cell $d_{tip}$. For transitioning a phalanx cell $P$ to a stalk cell $S$, we have to ensure that the distance from a given cell to a tip cell $T$ is less than a pre-defined threshold $d_{PS}$, while when transitioning back from a stalk cell $S$ to a phalanx cell $P$, we need to have a tip cell $T$ further than a pre-defined threshold $d_{SP}$. Finally, the transition from phalanx cell $P$ to tip cell $T$ ($P \rightarrow T$) is available only when there is no other tip cell in a certain immediate vicinity (specified by $d_{min}$) and the VEGF at the location of the cell exceeds a certain value $\alpha_v$. The distance to the nearest tip cell $d_{tip}$ is calculated based on the positioning grid, while the VEGF concentration is retrieved from the VEGF grid. The fixed thresholds used in the implementation can be found in Table \ref{tab:parameters}. These parameters are derived from baseline set of model parameter values in \cite{model_paper} after unifying both tumor and endothelial cell radii to 1 so as to make our model biologically sound.

\begin{table}[]
    \centering
    \caption{Table with the fixed parameters in our simulation}
    \begin{tabular}{p{0.08\textwidth}p{0.25\textwidth}p{0.03\textwidth}}
        Parameter    & Meaning & Value     \\
        $\alpha_v$   & VEGF threshold for $P \rightarrow T$ transition & 0.1  \\
        $\gamma$     & Fixed consumption rate & 10$h^{-1}$ \\
        $d_{SP}$     & Minimum distance from tip cell for $S \rightarrow P$ transition & 1.55 \\
        $d_{PS}$     & Maximum distance from tip cell for $P \rightarrow S$ transition  & 1.55 \\
        $d_{min}$   & Minimum distance from tip cell for $P \rightarrow T$ 
        transition & 10 \\
        $R$          & Endothelial cell radius  & 1    \\
        $S\_delay$   & Stalk cell may divide after this delay   & 4    \\
        $SS\_delay$  & Stalk cell growth time   & 3    \\
    \end{tabular}
    \label{tab:parameters}
\end{table}

In order to prevent transitions in conflict situation during simulation. The order of firing the transitions is set based on biological reasoning. We first fire the $S \rightarrow SS$ transition, as there is no notion in our biological framework that inhibits a cell from dividing when the conditions for division are met. The same reasoning applies to the $SS \rightarrow S$ transition (which we set as the second transition to be fired) as when the cell reaches the volume of a full stalk cell, it should be immediately assigned to that phenotype. As the possible divisions of the stalk cells are already conducted, the transition of the remaining stalk cells $S$ to phalanx cells $P$ can be freely handled. The last two transitions $P \rightarrow T$ and $P \rightarrow S$ are executed in this order. Even if generally these two transitions are not concurrent, in case such a situation might occur (based on the set values of $d_{min}$ and $d_{PS}$), we would like to prioritize the transition to a tip cell $T$ when the VEGF condition is met. By defining a clear ordering, any concurrency problems are mitigated.

The final component of Petri nets that we need to discuss is the token. In our model, a token represents a specific cell, encapsulating various characteristics of the defined cell. The token's internal representation is \{\textit{x}, \textit{y}, $d_{tip}$, \textit{VEGF}, $\tau$\}. \textit{x} and \textit{y} define the position of the cell in the positioning matrix. $d_{tip}$ specifies the distance from the current cell to the nearest tip cell (as computed using the positioning matrix). \textit{VEGF} indicates the VEGF concentration retrieved from the VEGF grid at the cell position \textit{(x, y)}. $\tau$ stores the amount of time that a cell spends in a given place.

\begin{figure}
    \centering
    \includegraphics[width=0.5\textwidth]{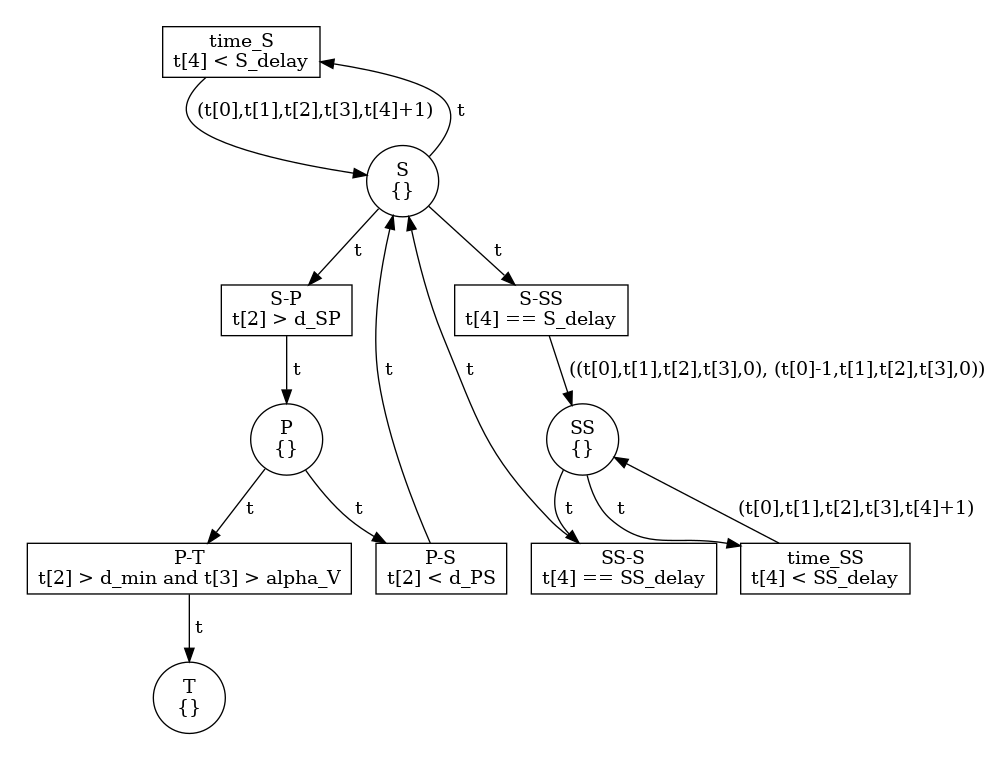}
    \caption{Schematic illustration of the Petri net.}
    \label{fig:petri_net}
\end{figure}

To create the Petri net, we used the Python library SNAKES \cite{Pommereau2015}, which provides flexibility in terms of modeling options or extending the features. The source code is available at {\href{https://github.com/LuLIACS/angiogenesis-modeling}{https://github.com/LuLIACS/angiogenesis-modeling}}. The net scheme can be seen in Figure \ref{fig:petri_net}. We can observe the places for each endothelial cell phenotype (marked by circles). There are two types of transition in the net. One type is a transition between different phenotypes. The other type of transition is a representation of time. The time is valid only for two places, \textit{S} and \textit{SS}. In all transitions, $t[i]$ refers to specific information preserved in the token at position $i$.

\subsection{Positioning and VEGF Grid} \label{sec:positioning_grid}
As previously introduced, there is a need in our model to determine the distance between two cells. To define a proper distance metric, one first needs to set the space in which the cells are placed. For convenience, we chose to model this environment as a 2-dimensional space, with the distance metric being the Euclidean distance. In this manner, the integrity of the biological concepts is preserved, and the extension to the 3-dimensional space is fairly straightforward. This kind of spatial representation can be interpreted as a cross section of the original environment.\\
The grid was implemented as a 2-dimensional NumPy \cite{harris2020array} matrix, with configurable size $(H, W)$. Each position in the matrix can hold a cell or zero value when it is empty. This introduces the limitation of all cells having the same volume, thus the same growth factor consumption rate $\gamma$. The cell is implemented as an object, initialized with the cell type (phalanx, stalk, or tip). Subsequently, the functions relevant to the cells are implemented on this matrix, such as cell movement (changing positions of cells), cell transition (changing cell types), calculation of the distance to the nearest tip cell, or division of stalk cells. All of these functions are connected to the Petri net and enabled for firing. In order to simulate the formation of new blood vessels, the matrix is initialized with all positions empty, except the bottom row, filled with phalanx cells $P$. This row is meant to depict a section of the micro-vessel culture in the microfluidic well-plate.\\

The VEGF grid is similar to the positioning grid and must match its shape. The VEGF grid was also implemented as a 2-dimensional NumPy matrix, which retains the concentration of VEGF present at each position. The VEGF grid is initialized with only the first row holding non-zero concentrations. It simulates the existence of some tumor cells that emit the VEGF signal. At each time step, a diffusion model is applied. The diffusion model should define the way in which the VEGF vector diffuses in our system. We implemented two types of diffusion. The first one is regular. The part of the VEGF that is diffused from the central position to a neighbor position with a lower value of VEGF is given by diffusion factor $f_d$, which is set to 0.1 in our implementation. Therefore, the value of VEGF in the center position and VEGF in the neighborhood position can be given by
\begin{equation}
VEGF_{center} = VEGF_{center} - f_d \cdot  VEGF_{center},
\end{equation}
\begin{equation}
VEGF_{neighbor} = VEGF_{neighbor} + f_d \cdot  VEGF_{center}.
\end{equation}
But they are too regular and pleasant for a natural phenomenon. Therefore, we implemented a second type of diffusion with some level of randomness. In this type of diffusion for each position in the grid, we randomly decide which part of the VEGF amount should diffuse in which direction \cite{random_diffusion}. We again look at each element in the VEGF matrix, and from this center element, a random part of the VEGF amount is diffused into each neighborhood position.\\
Additionally, the cells continuously consume part of the VEGF according to their consumption rate $\gamma$. In this manner, the VEGF grid has to be continuously updated.

\subsection{Cell Movement} \label{sec:cell_movement}
Regarding the cell movement, we adopt a simplified model. The movement is conditioned to happen only upwards. After a cell transition to a tip cell, if the tip cell is surrounded by stalk cells, the tip cell is allowed to update its position. An additional type of movement is the one when a stalk cell $S$ divides. The old cell has to be replaced by two new tokens which represent the division into two smaller cells. In order to create space for the two tokens, all of the cells positioned in the previous rows are simply shifted one row above, if there is space. In case there is no space, the cell division does not take place, and the movement does not happen. It particularly helps us to have a natural end of the simulation, instead of allowing cells to continuously multiply and push others outside the grid. This design choice can be easily adapted as needed.

\subsection{Time Integration}
\label{time_integration}
As illustrated in Figure \ref{fig:transitions_diagram}, certain transitions of the cell type are time constrained. Moreover, the cell movement and the diffusion of the VEGF have to be discretized in order to be integrated into the model. A natural way of doing this was to use time as our sampling factor and update both the cell positions and the VEGF concentrations once per time step. Thus, the time factor is a necessary piece in all components of our model.
We introduced a custom way of modeling the time in order to better match the biological scenario. In our approach, each cell (token) has an internal individual timer $\tau$ which is reset to 0 when the token enters a new place. This timer increases for each cell independently until it reaches the time constraint necessary for the connected transitions (i.e. for the $S \rightarrow SS$, the condition is $\tau > \tau_{sp}$). When the condition is met, the transition can be fired, and the token has its timer reset. The timer increase can be easily integrated into the Petri net using a loop transition so as to increment the $\tau$ variable inside the token representation, until a set threshold (i.e. $\tau_{sp}$). This loop transition should be fired at the start of each time step. In this manner, the biological model is not compromised when implementing the time aspect, as any number of cells are able to transition at the same time step if needed.

\section{\uppercase{Experiments and results}}
\label{Analysis}
\subsection{Diffusion simulations} \label{sec:simulations}
\begin{figure}[]
    \centering
    \subfigure[Cell positioning - Iteration 17]{
    \includegraphics[width=0.45\textwidth]{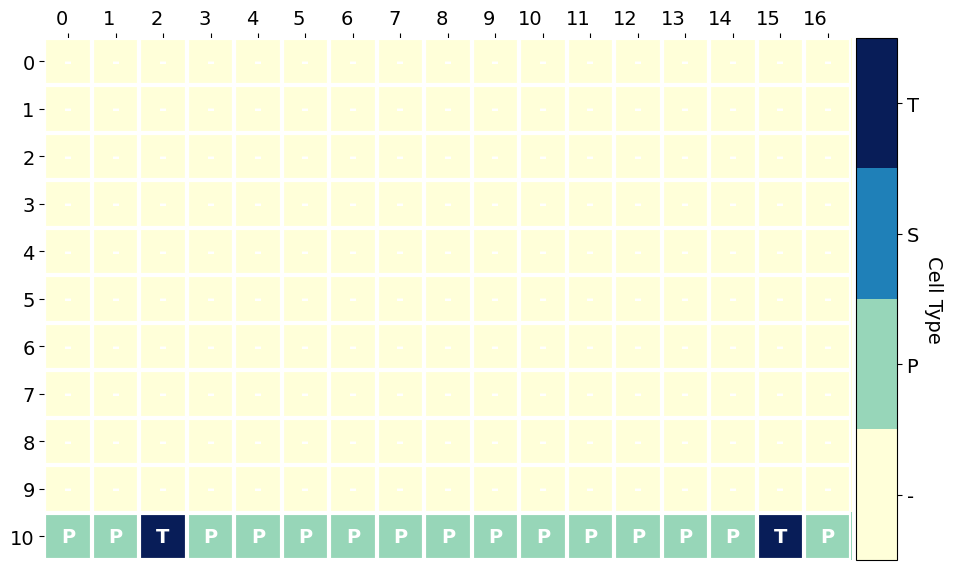}
    }
    \subfigure[Cell positioning - Iteration 92]{
    \includegraphics[width=0.45\textwidth]{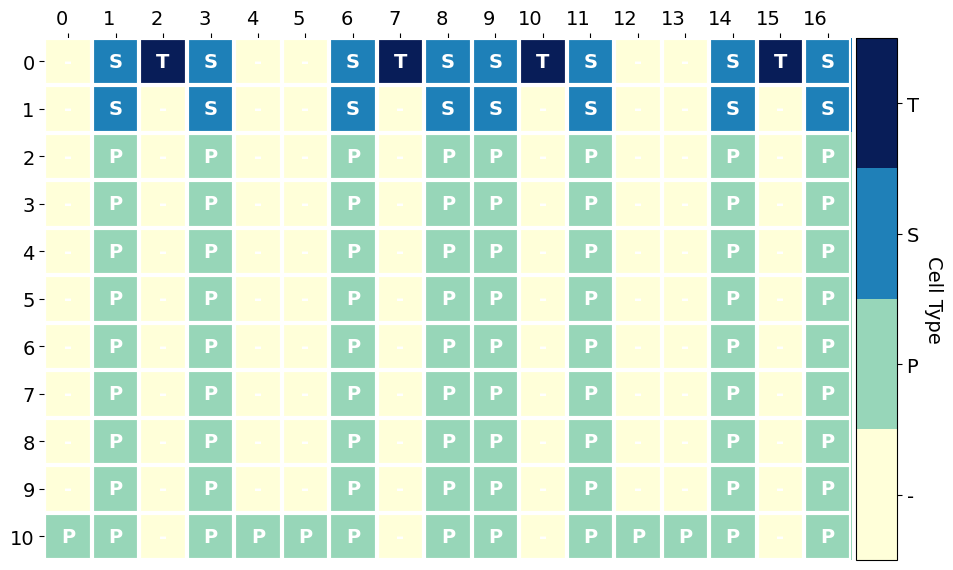}
    }
    \caption{The cell matrix for the first time step where tip cells are formed (left) and where the final movement of the tip cells is seen (right). The time step is shown below the figures.}
    \label{fig:normal_diffusion}
\end{figure}

\begin{figure}[]
    \centering
    \subfigure[Cell positioning - Iteration 39]{
    \includegraphics[width=0.45\textwidth]{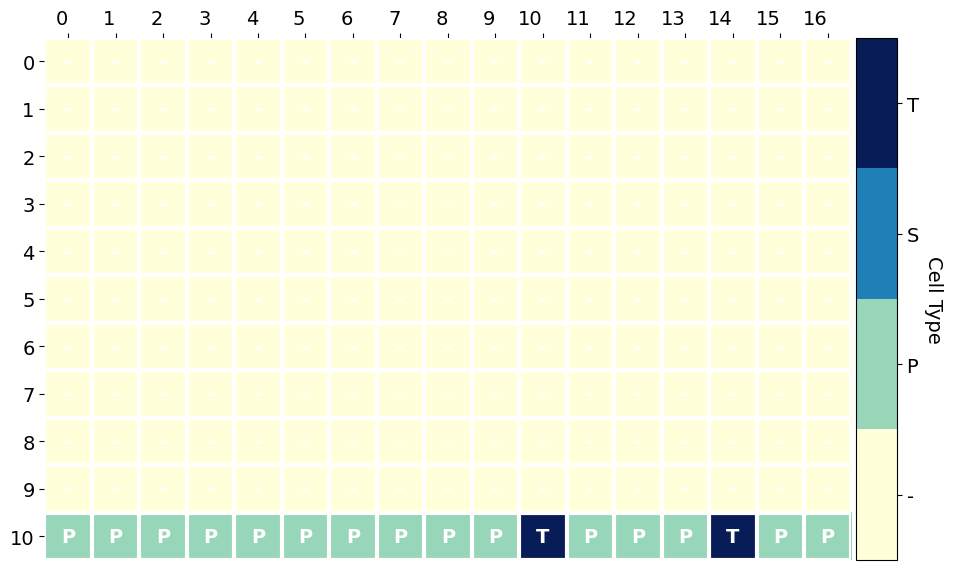}
    }
    \subfigure[Cell positioning - Iteration 136]{
    \includegraphics[width=0.45\textwidth]{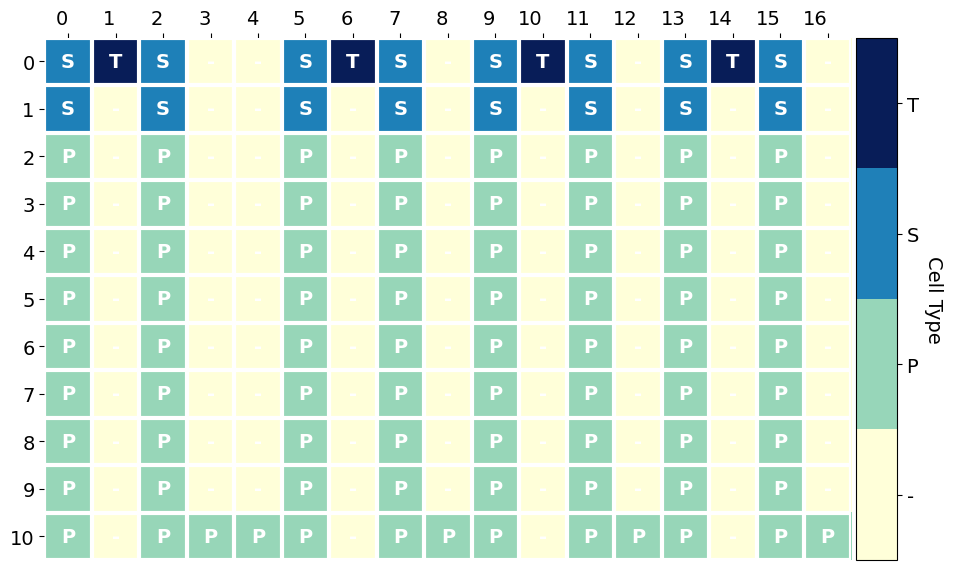}
    }
    \caption{The cell matrix during the random diffusion experiment when the first tip cells form (left) and when the final movement is made by a tip cell (right). The number below indicates the time step during the event.}
    \label{fig:random_diffusion}
\end{figure}

In the first experiment, the two diffusion types are compared with the same matrix initialization. The results of the simulations are shown in Figures \ref{fig:normal_diffusion} and \ref{fig:random_diffusion} for normal and random diffusion, respectively.

\begin{figure}[]
    \centering
    \subfigure[]{
    \includegraphics[width=0.45\textwidth]{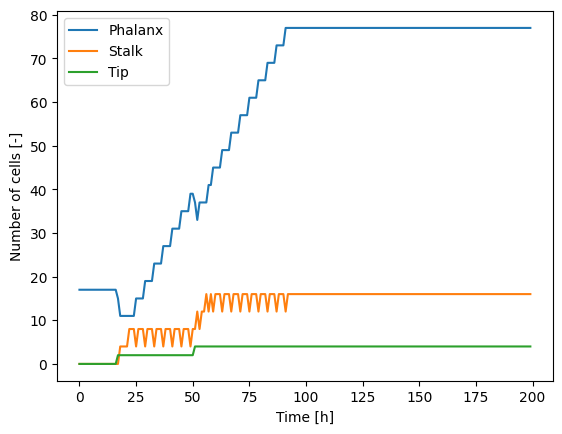}
    \label{fig:normal_diffusion_cell_count}
    }
    \subfigure[]{
    \includegraphics[width=0.45\textwidth]{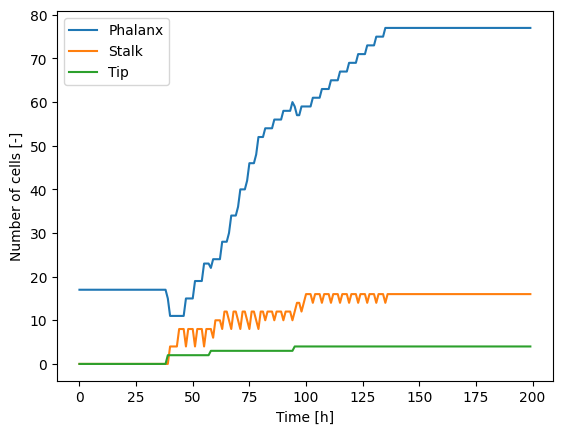}
    \label{fig:random_diffusion_cell_count}
    } 
    \caption{The number of cells shown per type for the normal and random diffusion experiments.}
    
\end{figure}

In the first simulation with the normal diffusion, tip cells form on two sides of the grid at the same time as shown in (a) of Figure \ref{fig:normal_diffusion}. When these tip cells are 1 square away from the top of the grid, enough distance has accumulated and the two new tip cells in the center form, both in time step 51. \\
The number of cells throughout the simulation is depicted for each cell type in Figure \ref{fig:normal_diffusion_cell_count}. In this graph the moments where the number of tip cells increases, followed by the constant addition of phalanx cells are visible.

The simulation with random diffusion, as shown in Figure \ref{fig:random_diffusion}, also has two phalanx cells that transfer to tip cells but on one side of the matrix at first, which happens 22 time steps later than with the normal diffusion. \\
The count of the different cell types for this experiment is shown in Figure \ref{fig:random_diffusion_cell_count}. The third transition to a tip cell happens on the left side at time step 58. The two initial tip cells reach the top at 80, so the increase in phalanx cells turns less steep. The fourth and final transition takes place as the third tip cell reaches the top after which the increase in phalanx cells is still less steep than during the growth of the first two branches. 

\begin{figure}[]
    \centering
    \subfigure[VEGF grid - Iteration 17]{
    \includegraphics[width=0.45\textwidth]{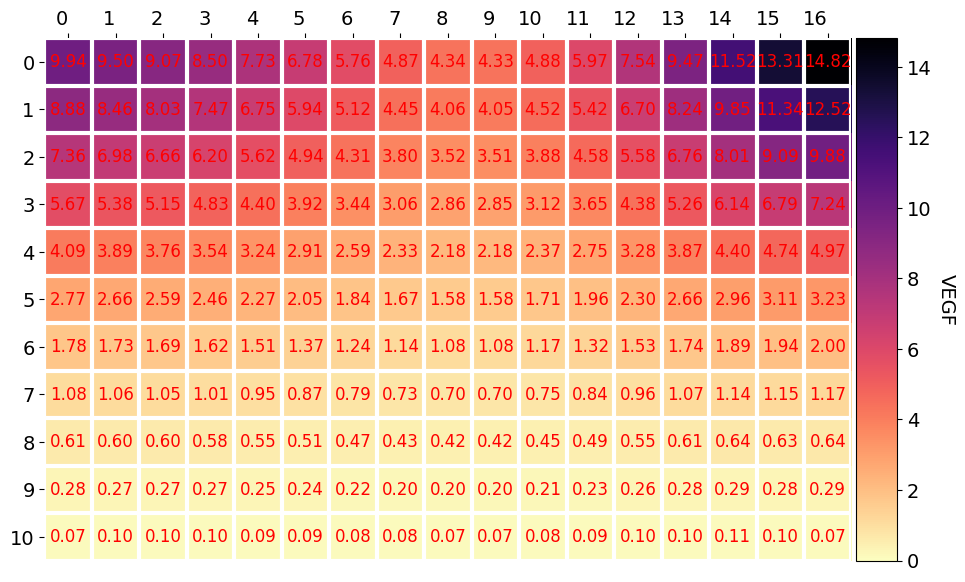}
    }
    \subfigure[VEGF grid - Iteration 92]{
    \includegraphics[width=0.45\textwidth]{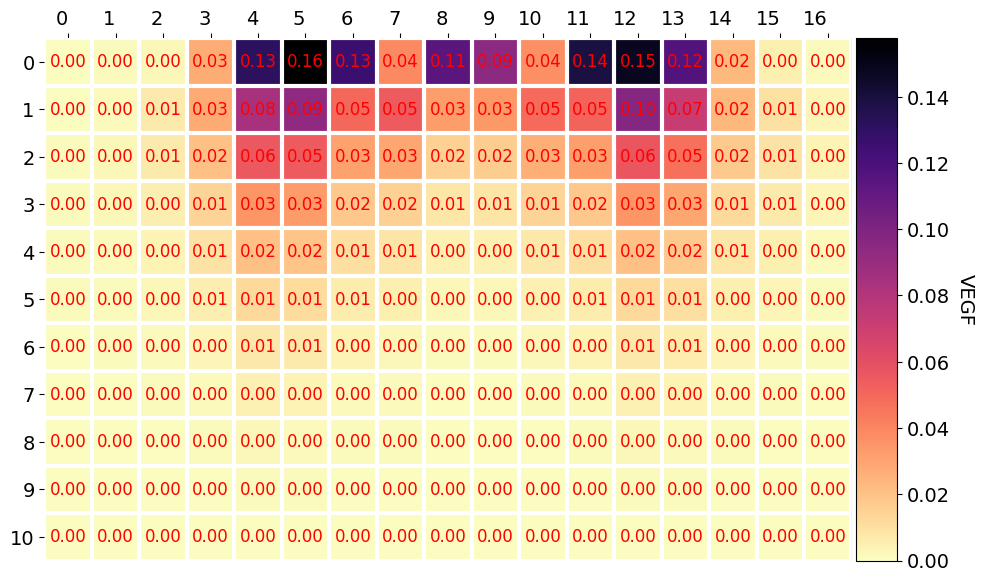}
    }
    \caption{The VEGF concentrations for the normal diffusion experiment, where the left frame shows the matrix when the first tip cell has formed and the right shows the matrix when the last tip cell reaches the top of the matrix.}
    \label{fig:normal_diffusion_VEGF}
\end{figure}

\begin{figure}[]
    \centering
    \subfigure[VEGF grid - Iteration 17]{
    \includegraphics[width=0.45\textwidth]{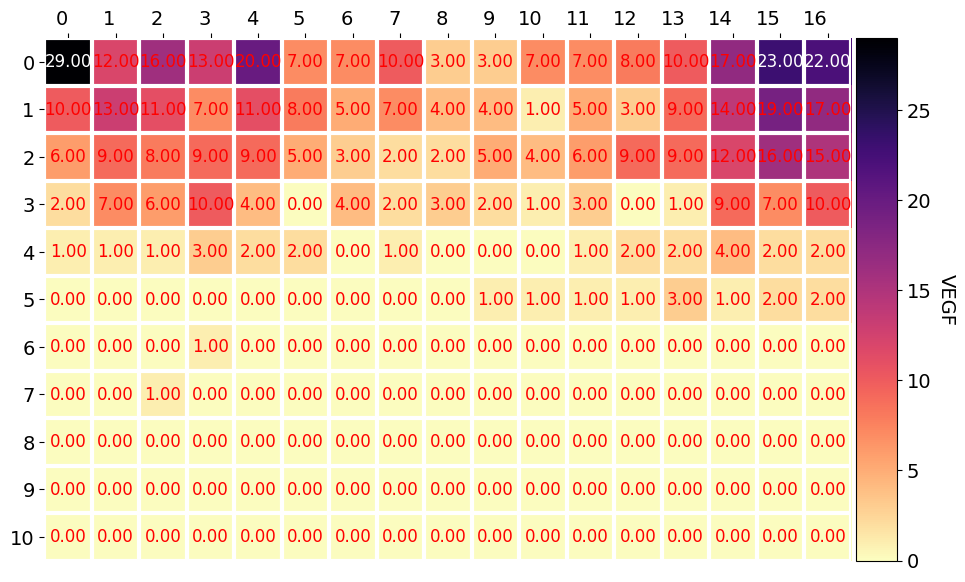}
    }
    \subfigure[VEGF grid - Iteration 92]{
    \includegraphics[width=0.45\textwidth]{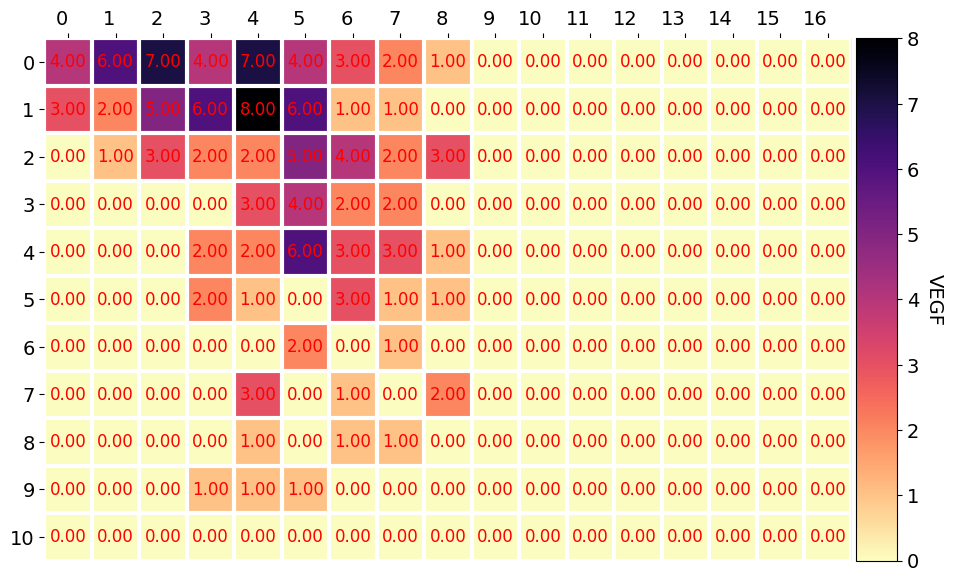}
    }
    \caption{The VEGF concentrations in the diffusion experiments at the same time steps as in the normal diffusion in Figure \ref{fig:normal_diffusion_VEGF}.}
    \label{fig:random_diffusion_VEGF}
\end{figure}

The corresponding VEGF matrices for the normal and random diffusion are shown in Figures \ref{fig:normal_diffusion_VEGF} and \ref{fig:random_diffusion_VEGF}. The states on the left show the concentrations at the time of the first transition to a tip cell, while the ones on the right are from the moment the last tip cell reaches the top of the grid. 
A structured gradient is observed and the matrix turns more symmetrical as time passes in the simulation. After 115 steps the matrix holds no VEGF anymore.

As random diffusion is not deterministic, different states could be reached with the same initial state. Note that the spreading patterns of the concentrations do look similar at the same phase despite using different diffusion methods. The key difference is the asymmetry that is observed, which follows from the branching that started at one side. 
We quantified the random diffusion, by running this experiment ten times and averaging as shown in Table \ref{tab:diffusion}.
\begin{table}[]
    \centering
        \caption{The average time step at which the described events took place, and the average number of branches are shown with their standard deviation.}
    \begin{tabular}{l|r}
    \textbf{Measurement} & \multicolumn{1}{c}{$\mu \pm \sigma$} \\ \hline \hline
    First tip cell & $33.6 \pm 6.7$ \\ \hline
    First grown branch & $74.6 \pm 6.7$ \\ \hline
    Number of branches  & $4.0 \pm 0.7$ \\ \hline
    Last grown branch &  $153.7 \pm 23.7$ 
    \end{tabular}
    \label{tab:diffusion}
\end{table}

The average number of branches is observed to be the same as in the normal diffusion simulation, which suggests that random diffusion can be used to approximate the normal diffusion.

\subsection{VEGF Amount Experiments}
\label{VEGF Amount Experiements}

In this set of experiments, we examined the influence of the initial value of VEGF that is inserted into our environment. From the biological point of view, we can assume that the environment closer to the tumor contains a higher level of VEGF concentration than the environment more distant from the tumor. Therefore, a small amount of VEGF inserted into the model may be seen as equivalent to the angiogenesis affected by the tumor in the distance. 

We performed these experiments with the same initial settings; only the initial VEGF value differed. We used the grid size $11 \times 15$ so that the cells have enough space to grow. Parameter $d_{tip}$ was set to 10. And then, we performed 200 iterations with each value of VEGF inserted into the grid. And all experiments were conducted for both types of diffusion. 

The results from the experiments are noted in  \ref{tab:vegf_random}, where we summed up the essential findings. \ref{tab:vegf_random} shows the results from the experiments with random diffusion and regular diffusion. The first column ("VEGF") shows the amount of initial VEGF. The second column ("First transition") indicates at which iteration the first transition of the cell phenotype occurred. The third column ("First branch") tells at which iteration the first branch was fully grown. The last column ("Number of branches") shows the number of branches that was grown in total. If the value is noted by '-', the situation didn't occur at all. 
\begin{table}[]
    \centering
    \caption{Table of results from experiments with different initial VEGF values, with random diffusion and regular diffusion.} 
    \begin{tabular}{|c|c|c|c|c|c|c|}
        VEGF  & \multicolumn{2}{c|}{FT}  & \multicolumn{2}{c|}{FB} &  \multicolumn{2}{c|}{\# branches}    \\
     	    & 	rand	& 	regu	  &  rand	& 	regu  & 	rand	& regu	  \\\hline
        50	     & 	27	& 	-	  & 	68	& 	-  & 	3	& 		-	  \\
        100	     & 	30	& 	-	  & 	72	& 	-  & 	4	& 		-	  \\
	150	     & 	36	& 	-	  & 	78	& 	-  & 	3	& 		-	  \\
        200	     & 	53	& 	-	  & 	94	& 	-  & 	3	& 		-	  \\
        300	     & 	43	& 	19  & 	84	& 	60  & 	3	& 		1	  \\
        400	     & 	32	& 	15  & 	74	& 	56  & 	3	& 		3	  \\
        500	     & 	26	& 	13  & 	68	& 	54  & 	4	& 		3	  \\
        600	     & 	33	& 	12  & 	73	& 	54  & 	3	& 		3	  \\
        700	     & 	27	& 	11  & 	68	& 	52  & 	4	& 		3	  \\
        800	     & 	37	& 	10  & 	78	& 	52  & 	5	& 		3	  \\
        900	     & 	35	& 	10  & 	76	& 	52  & 	3	& 		4	  \\
        1000	 & 	33	& 	9  & 	74	& 	50  & 	3	& 		3	  \\
    
    \end{tabular}
    FT: First transition; FB: First branch; \# branches: number of branches
    \label{tab:vegf_random}
\end{table}
Regarding to regular diffusion, we observe that the minimum amount of VEGF from the phalanx cell to the first tip cell is 300. Otherwise, the amount of VEGF, that diffuses to the positions of the initial phalanx cells, is trivial. Because it is consumed in each iteration until there is no more VEGF left. We can also observe a relation between the amount of VEGF and iteration where the first transition occurs. After exceeding the limit amount necessary for the first transition, the iterations in which the first transition took place decrease. Furthermore, the iterations in which the first branch was fully grown decreased. On the other hand, with random diffusion, we can not observe such relations. The iterations in which the first transition took place are around 27-53. But no pattern can be observed. 

In addition, no pattern can be observed in a number of branches related to the amount of VEGF. For example, the values are again random around 3-5. With regular diffusion, it seems that the number of branches is increasing with the amount of VEGF, but it does not hold in all cases.

In general, we can observe that the number of branches created is higher using random diffusion. Moreover, the first transition occurs sooner with random diffusion. Therefore, we can conclude that using random diffusion gives the model higher chances to transition and grow new branches.

\subsection{Sensitivity Analysis}
Some of the parameters \ref{tab:parameters} that we used in our model such as $\alpha_v$, $\gamma$, $d_{SP}$ and $d_{PS}$ are carefully estimated in \cite{model_paper}. Some of them are fine-tuned to fit our model scale such as $d_{min}$, $R$, $S\_delay$ and $ss\_delay$. Nevertheless, we don't know whether iPSC-ECs behave the same during angiogenesis as endothelial cells derived from adult mice or human. Therefore, the parameters should be able to adjust and we are interested to learn how sensitive is the model to the choice of parameters. We selected three parameters, which have space to adjust in our model, to conduct a sensitivity analysis. We set a range of each parameter so that the cell transition is still happening within 200 iteration and the branch still grows in a reasonable manner. A reasonable manner means that there is only one tip cell on top of a branch while growing and there is no single column branch that can grow without a tip cell. With all the factors considered, we set the range for the three parameters as following:

\begin{table}[]
    \centering
    \caption{Parameter range for sensitivity analysis}
    \begin{tabular}{p{0.08\textwidth}p{0.25\textwidth}}
        Parameter     & range     \\
        $\alpha_v$   & [0, 3]  \\
        $d_{SP}$     & [0.5, 2.5] \\
        $d_{min}$   & [10, 15] \\
    \end{tabular}
    \label{tab:parameters_range}
\end{table}

\begin{figure}[]
    \centering
    \subfigure[sensitivity indices for number of branches]{
    \includegraphics[width=0.45\textwidth]{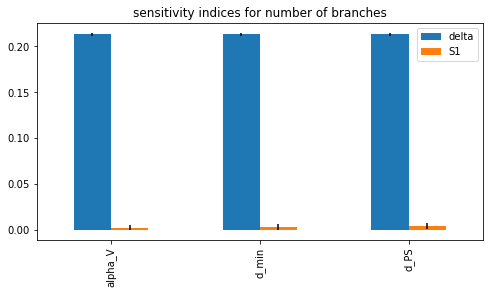}
    }
    \subfigure[sensitivity indices for first tip cell]{
    \includegraphics[width=0.45\textwidth]{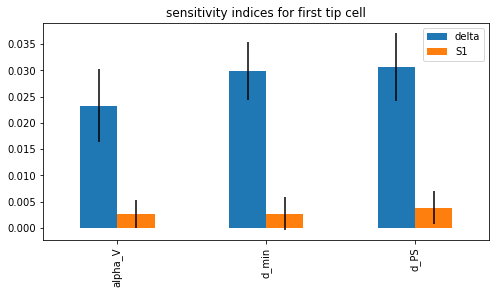}
    }
    \subfigure[sensitivity indices for last branch]{
    \includegraphics[width=0.45\textwidth]{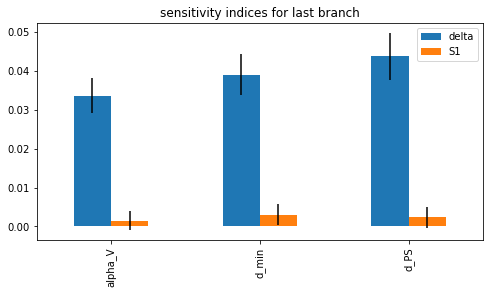}
    }
    \caption{Results of sensitivity analysis}
    \label{fig:sensitivity_anaysis}
\end{figure}

We selected three features to be the output of the model. They are the time step for the first tip cell, the time step for the last grown branch and the total number of branches. Delta Moment-Independent Measure \cite{Borgonovo2007} is used for sensitive analysis. It is a global sensitive indicator which looks at the influence of input uncertainty on the entire output distribution. In addition, the method returns first-order sobol indices which is a variance-based sensitivity  \cite{Sobol2009}. We set number of samples equal to 1024 per parameter. The sensitive indices are shown in Figure \ref{fig:sensitivity_anaysis}. We observed that, apart from the sensitive indices for number of branches, the rest of the indices are all very low and close to 0. That means the parameters do not contribute much to the uncertainty of the output of the model. That is probably because our analysis model is relatively small with a grid size $11 \times 15$. It does not give much variance in the output features. In the future, a larger grid size should be considered for sensitivity analysis.

\subsection{Scalability Experiment}
The final experiment that we conducted was designed to test the scalability of our current implementation.
To attain this goal, we ran our simulation for a logarithmically increasing grid size, with the height and weight varying in the interval $[10, 10\,000]$.
The current implementation has an exponential increase of running time, when the grid size is logarithmically increased.
Functions, such as the ones computing the distance to the closest tip cell, executing the cell division in the positioning grid and both VEGF diffusion mechanisms, are computationally heavy. In the future work, we will look into the possibility of improving these functions with the use of multi-processing or by vectorizing the computations.

\section{Conclusion}

In this paper, we modeled the transformations that iPSC-ECs undergo in the process of tumor angiogenesis in a microfluidics environment. The angiogenesis is guided by a gradient of VEGF. The model's behaviour was tested in various scenarios. Moreover, sensitivity analysis and scalability analysis are conducted to evaluate its performance.

The diffusion models were varied to find the most realistic one for model simulation. The diffusion approximation through random particle movement reached results that were very similar to those with normal diffusion. It suggests that random diffusion can be used to approximate the VEGF diffusion. The non-symmetric formation of branches is also more comparable to the chaotic branching seen in the \textit{in vitro} models. However, both approaches have their drawbacks. Where the normal diffusion is too perfect to be realistic, the random diffusion may result in too large `jumps' of particles to be realistic. In addition, we found that the initial amount of VEGF does not influence the speed of angiogenesis when using random diffusion. But it is an essential parameter for regular diffusion, where, with too low values of initial VEGF, the angiogenesis does not start at all.

In this work, we used a rather simplified model for cell movement, which only allows cells to move upward. Side-way movement of tip cells should be included in the future to make the model more realistic. Furthermore, the direction of motion should be decided by the concentration of VEGF, while preserving contact with stalk cells. 

Another limitation is the scale of the model with a grid size 11x15. It is relatively small to indicate the contribution of parameters to the uncertainty of the output of the model. A larger grid size is preferred for the sensitivity analysis. However, in scalability analysis, we observed that the current implementation has an exponential increase of running time when the grid size grows logarithmically. In the future work, functions with heavy computations should be further investigated and the limitation on scalability of the model should be solved using parallel computing.

\bibliographystyle{apalike}
{\small
\bibliography{references}}

\end{document}